# Mesoscopic structures and 2D hole systems in fully field effect controlled heterostructures

R.L. Willett, M.J. Manfra, L.N. Pfeiffer, K.W. West
Bell Laboratories, Lucent Technologies

ABSTRACT:
	Two fundamental extensions to the function of previously described fully field effect two-dimensional (2D) electron heterostructures are presented: First, using the same basic heterostructure design of lithographically defined contacts overlain by both an insulating layer and top-gate employed for *electron* systems, appropriate contact material allows a high mobility 2D *hole* layer to be populated. Second, a simple method for producing mesoscopic structures in these devices is presented in which small-scale metallic patterns are placed on the heterostructure under the insulating and global gate layers which allows local carrier density tuning via the overlapping gate arrays. Example devices using these generally applicable methods are demonstrated.

	In recent work [1] it has been shown that fully field effect 2D electronic heterostructures can be constructed with high mobilities, demonstrating extensive correlation effects.  The basic device design presented there is the starting point for elaboration to two fundamental modifications: preparation of high mobility hole systems and fabrication of fully field effect mesoscopic devices.
	Two-dimensional hole systems are an important component in the set of tools to study charge interaction effects due to the large effective mass and the varied spin-orbit coupling with respect to that of the electronic systems [2].  The ability to continuously tune the carrier density of a lower dimensional charge system enhances its utility. In this letter we describe fabrication of 2D hole channel fully field effect heterostructures, also displaying high mobility carrier transport with tunable densities. These devices are simpler in fabrication and design than previously detailed hole systems [3,4].
	 Myriad mesoscopic devices are produced using surface gates on or etching of heterostructures containing two-dimensional electron systems [5].  These devices have provided significant experimental access to lower dimensional electron physics [6,7]. An important aspect of operation in these devices is the ability to both locally and globally modulate the electron density, which allows probing of small-scale electronic properties and control of the Fermi sea to which the small scale electron population is linked. Presented here is a type of device with its central advantage local and global density control, achieved using a multilayer gate configuration fundamentally designed from fully field-effect heterostructures [1].  The general construction concept of the multilayer gate structure is described, as well an example of a functioning small scale conduction channel.

	Both the mesoscopic devices and p-type carrier systems utilize the undoped field–effect heterostructures whose construction has been previously presented [1]. To review pertinent aspects of this technique, a mesa is etched in a single interface AlGaAs/GaAs heterostructure sample, lithographically defined contacts are diffused to the AlGaAs/GaAs interface, an insulating layer of $Si_3N_4$ is deposited on the area of the mesa and contacts, and an overlay gate structure is then deposited over both the contacts and the mesa. For the electronic systems, upon positive bias of the top gate, electrons populate the AlGaAs/GaAs interface.

To produce the two-dimensional hole layer, the AlGaAs/GaAs heterostructure design is the same as that used for the electronic channel field effect devices. A schematic of the heterostructure layering dimensions and contacting structures used to produce hole channels are shown in Figure 1. On a (100) GaAs substrate an epitaxial GaAs layer is grown (typically 200nm), followed by $Al_xGa_{1-x}As$ forming the channel interface. In the results shown here the Al fraction x is .24. The $Al_xGa_{1-x}As$ layer is 100nm thick, capped by 5nm GaAs. A square mesa 500μm x 500μm is then etched onto the sample with wet etching to a depth of 200nm with $H_2O/NH_4OH/H_2O_2$, 100/10/2 by volume, 30% $H_2O_2$.

Into the perimeter of the mesa metal contacts are lithographically defined, evaporated, and diffused. Two metal layering schemes were performed and tested in these studies: 1) 250nm AuBe (Be~2%) on the heterostructure surface, followed by 60nm Au, and 2) 80nm AuBe (Be~2%), 50nm Ti, then 200nm Au. Both layering sets were then diffused in for ~ 15 minutes at 440°C. Following local protection of the contact pads away from the mesa, an insulating layer of amorphous $Si_3N_4$ is then deposited (standard PECVD deposition using a gas source reaction chamber) over the entire sample; layer thickness used here is 120nm. The final device preparation step is photolithographic definition of a top-gate that overlays the mesa and contacts as they themselves overlap the mesa (see Figure 1). The top gate is 30nm of aluminum, thermally evaporated at a rate of ~0.5nm/sec for 10 seconds, then >1.5nm/sec to achieve the target thickness. The AuBe/Au contact scheme has proven to be superior in contacting efficacy, with results described below.

Transport was measured in a four terminal configuration in a He3 refrigeration system. During cooling all sample leads are grounded and at low temperatures (~300mK) 2D layer population is achieved by negatively biasing the top overlay gate while maintaining all contacts at ground. Transport is then measured by standard lock-in methods where nominally one contact is sustained as ground reference during all current-voltage lead assignments.

Raw magneto-transport data is displayed in Figure 1. The Shubnikov-deHaas oscillations and the quantum Hall effect are resolvable but with a small resistive background. In spite of the inherent lower mobility of the hole system, the quantum Hall effect minima are substantial and fractional quantum Hall effect is observable in the range of filling factor one to two. The measured mobilities in these devices are commensurate with the observation of fractional quantum Hall features. In the device of Figure 1b the zero-field mobility is found to be ~ $9 \times 10^5 cm^2/V$-sec. The effect of gate voltage on carrier density is displayed in Figure 1c. With a population onset voltage of about 0.8V, the carrier density increases linearly up to an apparent saturation level of ~ $2 \times 10^{11} cm^{-2}$ in this $Al_xGa_{1-x}As$/GaAs sample with x = .24.

The density dependence of the gate voltage presents a simple operation mechanism related to that displayed by the electronic field effect heterostructure counterpart [1]. The onset of 2D layer population at roughly half the GaAs bandgap is consistent with mid-gap Fermi level pinning, and a corresponding top-gate voltage necessary to form an interface hole energy level. The saturation can be speculated to be due to population of low mobility surface states given a barrier that is exceeded by the voltage at onset of saturation. This is again similar to the electron systems described previously.

Turning now to fabrication of mesoscopic devices, in the multilayered gate structure for electronic devices [1] the mesoscopic defining gate is introduced at an intermediate point in the fabrication; see Figure 2. After mesa formation and contact definition and diffusion, the mesoscopic gate can be defined and evaporated directly onto the heterostructure mesa surface, with the requirement that the contacts have already been diffused into the heterostructure. This step is followed by deposition of the $Si_3N_4$ insulating layer and finally by definition and deposition of the top overlay gate. A micrograph of a sample device is displayed in Figure 2a in addition to the schematic map. 50 nm Al or 90nm Ti have been used for the mesoscopic gate material. In an alternative process, a thin (50nm) layer $Si_3N_4$ can be deposited after contact diffusion but before mesoscopic gate construction in order to minimize leakage through incidental wafer defects of the mesoscopic gate to the 2D electron layer.

Results for devices with short path length top gates and narrow channels were tested for field-effect multilayer devices. The short path length top gate extends across the conducting path of the underlying 2D layer as shown in the Figure 2c inset schematic. Two-terminal transport data for a short path length (~0.5 μm) top gate in which the overlay top gate and the mesoscopic gate are both biased to +2V to populate the 2D layer are shown in Figure 2c. Following this loading of the AlGaAs/GaAs interface, the mesoscopic gate bias is reduced ultimately leading to full depletion of the 2D layer under that gate near 0.8V. In reaching turn-off of the conduction under the short-length gate, note the linear decrease in transmitted current for voltages from ~1.1V to 0.8V.

A narrow conducting *channel* in the 2D electron layer can be produced by defining a top gate as shown in the schematic inset of Figure 2d. Data are displayed for bulk conduction and conduction through a 2μm x 4μm channel defined by the top gate in which the overlay top and mesoscopic gates have applied gate voltages tuned to effect conduction. Four terminal magneto resistance dramatically demonstrates the effect of B=0 diffusive backscattering in the narrow channel, which is suppressed by application of a magnetic field [7]. Concomitantly the Shubnikov-deHaas oscillations indicate a change in the local electron density.

To summarize the findings, it is possible to produce a fully-field effect 2D hole system using a single interface AlGaAs/GaAs heterostructure with AuBe contacts and a simple overlapping surface gate structure. High mobility transport for a 2D hole system is displayed by such devices, with hole density modulation possible by tuning the top-gate bias. These devices hold promise for fundamental studies in lower dimensional hole systems and potentially for extension to electron-hole bilayer devices. Also using this basic heterostructure design but with electronic contacting, intermediate layering of mesoscopic gate structures can result in multiply tunable small-scale devices with respect to the local electron densities. These devices offer the potential for examining mesoscopic effects in high mobility structures not damaged by etching.

FIGURE CAPTIONS:
Figure 1. field effect hole device: top panel a) Schematic cross-section of p-type field effect heterostructure device. middle panel b) d.c. magneto-transport through a p-type field effect heterostructure device using AuBe/Au contacting as outlined in the text. This is longitudinal resistivity from a four terminal measurement showing properties of the quantum Hall effect. The density is roughly $1.8 \times 10^{11}$ cm$^{-2}$. The temperature is 280mK. Bottom panel c) 2D hole density as a function of gate voltage for p-type field effect heterostructure device using AuBe/Au contacting. This $Al_xGa_{1-x}As/GaAs$ sample has x = .24, and the measurements are made at 280mK.

Figure 2. field-effect mesoscopic device: top panel a) Schematic and photomicrograph of a mesoscopic gate structure (short-path length gate) embedded in a non-doped field effect device. The mesa edge length is 500μm, with contacts, mesoscopic gate structure and top gate structures overlaying the mesa as shown in the schematic. middle panel b) short path length mesoscopic gate schematic design and two terminal conductance as a function of mesoscopic gate voltage at 280mK. bottom panel c) narrow channel schematic design and magneto-resistance through the 2μm x 4μm channel (red) and for bulk (black). The gate voltages are adjusted to populate the channel, which demonstrates diffusive backscattered resistance which is suppressed upon application of a magnetic field [7]. Temperature is 280mK.

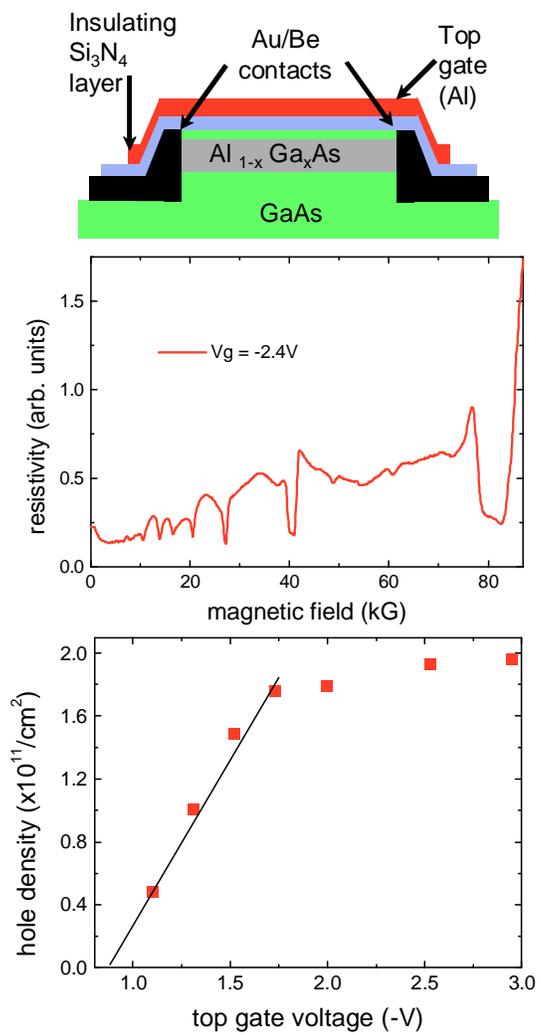

Figure 1

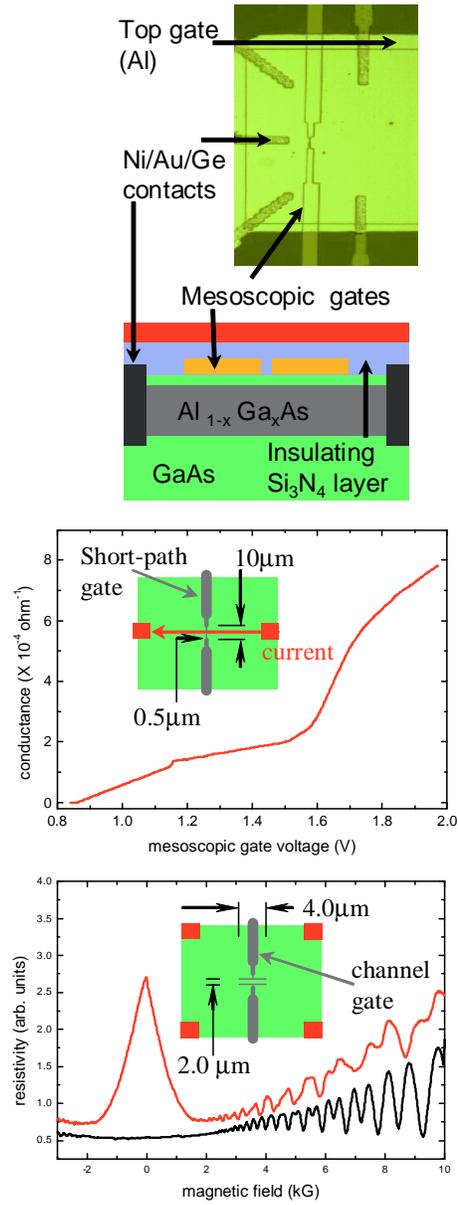

Figure 2